\documentclass[12pt]{amsart}
\usepackage[dvips]{color}
\usepackage{amsmath}
\usepackage{amsxtra}
\usepackage{amscd}
\usepackage{amsthm}
\usepackage{amsfonts}
\usepackage{amssymb}
\usepackage{eucal}
\usepackage{epsfig}
\usepackage{graphics}
\def\({\left(}
\def\){\right)}

\newcommand{\mub}{\mbox{\boldmath$\mu$}}

\newcommand{\betab}{\mbox{\boldmath$\beta$}}
\newcommand{\gammab}{\mbox{\boldmath$\gamma$}}

\newcommand{\nn}{\nonumber}
\newcommand{\bea}{\begin{eqnarray}}
\newcommand{\ena}{\end{eqnarray}}
\def\bel{\begin{eqnarray}}
\def\enl{\end{eqnarray}}
\newcommand{\be}{\begin{eqnarray*}}
\newcommand{\en}{\end{eqnarray*}}
\newcommand{\ba}{\begin{array}}
\newcommand{\ea}{\end{array}}

\newcommand{\slth}{\widehat{\mathfrak{sl}}_2}

\newcommand{\Tr}{{\rm Tr}}

\newenvironment{tenumerate}{
  \begin{enumerate}
  
  }{\end{enumerate}}
\newcommand{\bi}{\begin{tenumerate}}
\newcommand{\ei}{\end{tenumerate}}
\newcommand{\isoto}[1][]%
{{\mathop{\buildrel{\sim}\over\longrightarrow}\limits_{#1}}}

\def\[{\left[}
\def\]{\right]}
\newcommand{\la}{\lambda}

\newcommand{\al}{\alpha}

\newcommand{\z}{\zeta}

\numberwithin{equation}{section}

\def\bi{\mathbf{i}}

\def\Io{I_\mathrm{odd}(m)}

\begin{document}

\begin{title}{On one-point functions for sinh-Gordon model at finite temperature.}

\end{title}
\author{S.~Negro and  F.~Smirnov}

\address{SN
Laboratoire de Physique Th{\'e}orique et
Hautes Energies, Universit{\'e} Pierre et Marie Curie,
Tour 13, 4$^{\rm er}$ {\'e}tage, 4 Place Jussieu
75252 Paris Cedex 05, France
\newline
\centerline{$\mathrm{and}$}
\newline
\hspace{0.3cm}Dipartimento di Fisica, Universit\'a degli Studi di Torino
Open Space Dottorandi, via Pietro Giuria 1, 10124, Turin}\email{negro@to.infn.it \hspace{0.1cm};\hspace{0.1cm} negro@lpthe.jussieu.fr}

\address{
FS\footnote
{Membre du CNRS}Laboratoire de Physique Th{\'e}orique et
Hautes Energies, Universit{\'e} Pierre et Marie Curie,
Tour 13, 4$^{\rm er}$ {\'e}tage, 4 Place Jussieu
75252 Paris Cedex 05, France}\email{smirnov@lpthe.jussieu.fr}

\begin{abstract}
Using fermionic basis we conjecture the exact formulae for the expectation
values of local fields in sinh-Gordon model. The conjecture is checked against
previously known results. 
 \end{abstract}

\maketitle

\section{Introduction}
The importance of the one-point functions for study of ultra-violet asymptotics of
the multi-point correlation functions is nicely explained in \cite{Alpert} (see also \cite{FFLZZ}).
This explanation was repeated twice in \cite{OP,HGSV}, so, we would not return to it here.
In the present paper we consider one-point functions for the sinh-Gordon (shG) model whose Euclidean action
is given by
\begin{align}
\mathcal{A}&=
\int \left\{ \Bigl[\frac 1 {4  \pi} \partial _z\varphi (z,\bar{z})\partial_{\bar{z}}\varphi (z,\bar{z})+
\frac{2\mub ^2}{\sin\pi b^2}\cosh(b\varphi(z,\bar{z}))
\right\}\frac{idz\wedge d\bar{z}}2\,.
\label{action1}
\end{align}
For the shG model $0<b^2<\infty$, but we shall often compare it with the sine-Gordon model (sG)
for which $-1\le b^2<0$. The shG model obeys the duality $b\to 1/b$, so, one may restrict
consideration to $0<b^2\le 1$. 

We put the model on the cylinder of circumference $2\pi R$. 
The generatrix 
and the directrix of the cylinder
will be called space and Matsubara directions respectively.

We consider the expectation
values of the 
primary fields $e^{a\varphi(0)}$ and their descendants
which we denote collectively $\mathcal{O}_a$. 
There are two  ways to define the descendants
in agreement with the following picture which we adopt with
all possible reservations. 

One
way is  take for $\mathcal{O}_a$
the normal ordered product
(derived from the T-ordering with respect to the Euclidean time
$\log(z\bar z)/2$)  of $e^{a\varphi(0)}$ with any polynomial in derivatives
of $\varphi(0)$. 
We
can  consider shG model as a perturbation of
$c=1$ CFT, so we call these descendants the Heisenberg descendants.
We take only even degrees polynomials which is not a real restriction
as will be explained soon. Another way 
consists in defining the modified energy-momentum 
tensor, the one whose trace is proportional to $e^{-b\varphi (z,\bar z)}$, and to  
take 
 for $\mathcal{O}_a$ the normal
ordered products with derivatives $\partial_z^kT_{z,z}(0)$, $\partial_{\bar{z}}^kT_{\bar z,\bar z}(0)$
We restrict ourselves to considering only 
the derivatives of even order. We
call these descendants the Virasoro descendants 
considering the shG model as a perturbation
of the Liouville model by a primary field $e^{-b\varphi (z,\bar z)}$. 
This gives rise to reflection relations as explained below.

An attempt to follow the picture above literally 
leads to IR divergent integrals. 
The divergencies cannot 
be cured by putting the theory on a cylinder since 
the scaling dimensions of perturbations are negative.
On the other hand
our final goal is the reflection relations which reflect the UV structure
of the theory and, hence, should not really depend on the regularisation.
As explained in \cite{FLZZ2} the IR problem can be treated introducing 
a metric on the world sheet. In any case, justification of reflection relations goes far
beyond the scope of the present paper. We rather take them for granted and
investigate their consequences verifying  agreement with previously known results. 

The shG model provides a simplest example of integrable quantum field theory.
That is why the problem of computing the finite
temperature one-point functions attracted
attention before. In the paper \cite{LM}  LeClair and Mussardo propose a method
to obtain series for the one-point function of primary fields, based on known
form factors. So, one can say that the present paper adds two new things: sums up the
series in the form of integral equation and allows to consider the descendants. Another interesting
work on the one-point functions of primary fields is due to Lukyanov \cite{luksep}, we shall comment
on it at the very end of the present paper. 

Consider the one-point function 
\begin{align}
\langle\mathcal{O}_a(0)\rangle  _R\,.\nn
\end{align}
The integrability of the shG model is based on the existence of an infinite number
of local integrals of motion. 
They act by commutators on the local operators. Obviously, the one-point functions
of the descendants obtained by this adjoint action  vanish, which explains the restrictions which
we imposed onto Heisenberg and Virasoro descendants. 
The idea of the paper \cite{FFLZZ} consists in the following. The one-point functions of
the Heisenberg descendants are symmetric with respect to
\begin{align}\sigma_1:\ a\to-a\,.\label{refl1}\end{align}
On the other hand it is natural to assume that the one-point functions of the
Virasoro descendants inherit the symmetry of the Liouville model:
\begin{align}\sigma_2:\ a\to Q-a\,.\label{refl2}\end{align}
In the conformal limit the two types of descendants are related by certain reflection
matrix. Since we refer for generic $a$ to one-to-one correspondence between the operators in
the massive model and its ultra-violet limit the one-point functions must satisfy
Riemann-Hilbert problem associated with the reflection matrix.
This Riemann-Hilbert problem is called the reflection relations in the paper.

The reflection relations themselves reflect the UV structure of the theory.
However, they are rather useless if the analytical properties of one-point
functions in $a$ are not known. 
These properties depend on the the IR environment, understanding them 
is a complicated issue.

Up to now the one-point functions were available in full generality
only for the sG model \cite{HGSV}. 
Obviously, the one-point functions on the cylinder are 
closely related to the ground state in the Matsubara direction. For the sG model
the ground state is complicated. This results in rather involved analytical properties
of one-point functions with respect to $a$. 
Namely,  for finite $R$,  the analytical continuation
of the one-point function from the region $0<a<Q$ does not possesses symmetry
with respect to both reflections, the symmetry under $\sigma _1$ is broken
being replaced by rather complicated analytical structure described in the paper \cite{SS}.
The situation changes in the plane
limit $R=\infty$ when symmetry under both
reflections $\sigma _1,\sigma _2$ holds. In principle they specify the one-point function
up to unknown even, $Q$-periodic multiplier. The latter can be fixed from some
minimality assumptions \cite{FLZZ1,FLZZ2,FFLZZ}. 

For shG model the Matsubara ground state is supposed 
to be much simpler than in the sG case,
that is why it is natural to assume that both symmetries take place for finite $R$.
This assumption agrees with the large $R$ expansion of \cite{LM}, and with the 
classical limit \cite{luksep}.
So, the $R$-dependent part as a function of $a$ is even and $Q$-periodic. This function
cannot be found from the reflection relations, only in the limit $R=\infty$ one can assume that
the one-point functions
coincide with the analytical continuation with respect to $b$ from the sG case. One
more problem with the reflection relations is that the Riemann-Hilbert problem is
hard to solve for the descendants on level 4 and higher even for the case $R=\infty$.

From what has been said one may get an impression that finally the reflection
relations are not very useful for computing the one-point functions at finite $R$. We would
like to emphasise that this impression is wrong, and the reflection relations
are actually very useful if one understands how to utilise them \cite{NS}.

For finite $R$ the one-point functions for the sG model were found 
using the fermionic structure of the space of local operators in \cite{HGSV}.
This paper relies heavily upon \cite{HGSIV} where the conformal case was considered. 
The computation of \cite{HGSV} may be not completely rigorous from
a mathematician's point of view, but it is sufficiently  reliable for a  physicist.
Indeed,  the fermionic structure used there is derived from the fermionic structure
of lattice model \cite{HGSII}, and the fact that the expectation values on the cylinder
are expressed as determinants, provided the fermionic basis is used, is a corollary
of a similar fact for the lattice model on the cylinder \cite{HGSIII}. So, we begin the
computation with the sG model with both ultraviolet and infrared cutoffs imposed. 
In many respects we shall proceed for the shG case by analogy with the sG one.

In order to make  the comparison with \cite{HGSV} easier
we shall use in what follows the notations:
\begin{align}
\nu =1+b^2, \qquad \al =\frac {2a}{b+b^{-1}}\,.
\end{align}
In terms of $\al$ our reflections are
$$\sigma _1\ : \ \ \al\to -\al \,,\qquad \sigma _2\ : \ \ \al\to 2-\al\,.$$
We shall also denote $e^{a\varphi (0)}$ by $\Phi _\al(0)$.

The main tool used in \cite{HGSV} is the fermionic basis. In CFT the descendants are created
by action of two Virasoro algebras. Passing to the massive theory one argues that, at
least for generic $a$, there is an unambiguous 
identification of local operators with CFT descendants. The
idea of \cite{OP,HGSV} is that 
when describing the integrable
deformation
it is convenient to switch from the usual basis of Verma module 
created by
Virasoro generators to the one created by fermions, two for
both chiralities: $\betab^*_{2j-1}$,  $\gammab^*_{2j-1}$, $\bar{\betab}^*_{2j-1}$, 
$\bar{\gammab}^*_{2j-1}$. It has been said that we are interested in the
space of descendants modulo action of the local integrals of motion. The
basis of the quotient space is given by
\begin{align}
\betab^* _{I^+}\bar{\betab} ^*_{\bar{I}^+}
\bar{\gammab }^*_{\bar{I}^-}\gammab^*_{I^-}
\Phi _{\al}(0)\,,
\quad
\#(I^+)=\#(I^-),\ \  \#(\bar I^+)=\#(\bar I^-)\,,
\label{basis}
\end{align}
here and later we use the multiindex notations:
\begin{align}
&I=\{2i_1-1,\cdots ,2i_n-1\}\,,\quad |I|=\sum_{p=1}^n(2i_p-1)\,,\nn\\&
 \betab^*_I=\betab _{2i_1-1}\cdots \betab _{2i_n-1}\,,\quad
\gammab^*_I=\gammab _{2i_n-1}\cdots \gammab _{2i_1-1}\,,\nn
\end{align}
and similarly for the second chirality. According to \cite{NS}, the fermionic basis 
is defined from the reflection properties:
\begin{align}
\sigma _1,\sigma _2\ :\ \betab^*_{2m-1}\ \leftrightarrow\ \gammab^*_{2m-1}\,.\label{reflections}
\end{align}
 The definition of fermions in this paper is changed
comparing to \cite{HGSIV,OP,HGSV} by a ``CDD multipliers" in order that they possess, in
addition to \eqref{reflections}, 
duality under $b\to1/b$:
\begin{align}
\mathrm{duality}\ :\ \betab^*_{2m-1}\ \leftrightarrow\ \gammab^*_{2m-1}\,,\label{duality}
\end{align}
as will be explained in Section \ref{fermions}.

The main breakthrough of the paper \cite{HGSV} comparing to \cite{HGSIV,OP}, where
the fermions were used only for creating descendants,  consists in finding
relations which allow to 
use fermions in order to
shift the primary operator. Namely,
let us relax the condition $\#(I^+)=\#(I^-)$,  $\#(\bar I^+)=\#(\bar I^-)$ keeping
$\#(I^+)+\#(\bar I^+)=\#(I^-)+\#(\bar I^-)$, and denote  $m=\#(I^+)-\#(I^-)$, 
assuming $m>0$ for the sake of definiteness, then
\begin{align}
&\betab^*_{I^+}\bar\betab^*_{\bar{I}^+}
\bar\gammab^*_{\bar{I}^-}\gammab^*_{I^-}
\Phi_{\alpha+2m\frac{1-\nu}{\nu}}(0)\label{idm+}\\
 &\cong 
 \frac  { C_m(\al)}{\prod_{j=1}^mt_{2j-1}(\al)}
\ \betab^*_{I^+ +2m}\bar\betab^*_{\bar{I}^+-2m}
\bar\gammab^*_{\bar{I}^- +2m}\gammab^*_{I^--2m}\betab^*_{\Io}\bar\gammab^*_{\Io}
\Phi_{\alpha}(0)\,,\nn
\end{align}
where  the sign $\cong$ means that the identification holds in a weak
sense (under the expectation value). The operators with negative indices, if any, are understood as annihilation operators:
\begin{align}
&\betab^*_{-(2j-1)}=\gammab_{2j-1}, \ \ \gammab^*_{-(2j-1)}=\betab_{2j-1},
\ \ \bar\betab^*_{-(2j-1)}=\bar\gammab_{2j-1}, \ \ \bar\gammab^*_{-(2j-1)}=\bar\betab_{2j-1}\,,\label{comm}\\
&[\betab_a,\betab^*_b]_+=-t_a(\al)\delta_{a,b},\ [\gammab _a,\gammab^*_b]_+=t_a(-\al)
\delta_{a,b}\,,\nn\\
&[\bar\betab_a,\bar\betab^*_b]_+=t_a(-\al)\delta_{a,b},\ [\bar\gammab _a,\bar\gammab^*_b]_+=-t_a(\al)
\delta_{a,b}\,.\nn
\end{align}
The constant $ C_m(\al)$ and $t_a(\al)$ will be given in Section \ref{fermions}.

Let us comment on the status of  relations \eqref{idm+}.
The paper \cite{HGSIV} where the conformal limit of the 
fermionic operators
is presented, originally defined for the lattice six-vertex model, is the starting point of all 
further development. The relations \eqref{idm+} were observed using the formulae of \cite{HGSIV} and
comparing them with the CFT three-point functions.
Then they were promoted to the sG model by identification of local operators.

In the present paper we shall use the relations \eqref{idm+} for the shG model.
Since the procedure used in the sG case is not available, the status of these 
relations is much more shaky. However, it should be possible to verify them
computing the UV $R\to 0$ limit of our final formulae and comparing it with the
Liouville three-point functions \cite{ZZ}. This has not been done, we hope to return
to the study of the UV limit in future. 

Provided the reflection relations hold in the shG model for finite $R$, 
the one-point functions in the fermionic basis must be
periodic with period 2 functions of $\al$. Here we make one more
assumption that exactly as in the sG case these functions are given by determinants
of matrices with the following matrix elements \footnote{ In \cite{HGSV} the spectral parameter $\z$ and Mellin transform are used. Here we use instead the ``self-dual" parameter $\theta=\log(\z)/\nu$ and the  Fourier transform. To compare formulae one has to
identify $k=2\nu k^{\mathrm{HGSV}}$. }
$$ \Theta\(
 ia, ib|\al\)
-\pi \mathrm{sgn}(a)\delta _{a,-b}  t_{a}(\al)\,,\quad a,b\in \mathbb{Z}/2\mathbb{Z}\,.$$
Assuming that it is not very hard to guess what the function $ \Theta\(
 l,m|\al\)$ should be from two requirements. First, the  one-point functions for the components of the
 energy-momentum tensor, which 
 can be computed
 from general reasons,  have to be reproduced. Second, the relations \eqref{idm+} give rise
 to certain compatibility conditions which result in a  rather restrictive equation for 
 $ \Theta\(
 l,m|\al\)$ ({\it cf.} (9.4) in \cite{HGSV}).  These two
 requirements allowed us to conjecture the form of the
 function $ \Theta\(
 l,m|\al\)$ in Section \ref{shG}.
 It satisfies the symmetry relations
 \begin{align}
& \Theta\(
k, l|-\al\)= \Theta\(
l,k|\al\)\,,\quad  \Theta\(
k, l|\al+2\)= \Theta\(
k, l|\al\)\,.\label{symthetashG2}
\end{align}

The formula \eqref{idm+} implies an infinite number of consistency
relations for $ \Theta\(
k, l|\al\)$ evaluated at odd integer imaginary points, we 
discuss this in Section \ref{shG}. Similarly to the sG case \cite{HGSV} all of
them follow from one identity \eqref{shiftshG}.

Using  \eqref{idm+} and the periodicity with period $2$
\eqref{symthetashG2} one obtains the one-point functions 
for the descendants of the primary fields with $\al =\al +2m+2n\frac{1-\nu}{\nu}$
in terms of $ \Theta\(
k, l|\al\)$. For irrational $\nu$ these points are dense in $\mathbb{R}_+$. 
This means that we could start, for example, from $\al=0$, and then obtain the one-point
functions for arbitrary $\al$ simply by continuity. This may be useful for justification of
our conjecture in view of the following interesting observation.  

The function $ \Theta\(
k, l|\al\)$ is defined starting from the TBA equation which is very simple in the shG case
(see Section \ref{shG}). 
Let us generalise the TBA  considering instead of 
the partition function of the Gibbs ensemble
$$Z(R)=\Tr \(e^{-2\pi RH}\)\,,$$
the partition function of the generalised Gibbs ensemble \cite{RDYO} 
$$Z(\{g_{2j-1}\})=\Tr \(e^{-\sum_{p-\infty}^{\infty}g_{2p-1}\mathcal{I}_{2p-1}}\)\,,$$
where for $j\ge 1$ the operators $\mathcal{I}_{2j-1}$, $\mathcal{I}_{-(2j-1)}=\bar{\mathcal{I}}_{2j-1}$ are local integrals of motion, $g_{2p-1}>0$. 
These space local integrals are obtained by integrating the corresponding densities along the
space axis, we use calligraphic letters  in order to distinguish them from 
the Matsubara ones.
Usual partition function of the Gibbs ensemble is obtained 
from the partition function of the generalised Gibbs ensemble by specialisation:
\begin{align} g_{2p-1}=0, \ |2p-1|>1\,; \quad g_{\pm 1}=2\pi R\,.\label{sub}\end{align}
We explain in Section \ref{further} that to the generalised Gibbs ensemble
one can associate a function $Y(\{g\})$,
which shall refer to as on-shell Yang-Yang action. 

Let us be precise with the terminology. 
The Yang-Yang action was introduced in \cite{YY} as a rather formal object
needed to prove the existence of solutions to the Bethe Ansatz equations for repulsive Bose gas. 
So, the Yang-Yang action depends upon the Bethe numbers $\la _j$ producing the 
Bethe equations when varied with respect to  $\la _j$. First unexpected application
was found by Gaudin \cite{Gaudin} who observed that the norm of any Bethe vector is
expressed as Hessian composed of the second derivatives
of the  Yang-Yang action with respect to $\la_j$. 

After many years of oblivion the Yang-Yang
action suddenly surfaced in the study of $N=2$ supersymmetric models \cite{GerSam,NikSam}.
Then it was used in \cite{lukYY} in a slightly different fashion. 
Our understanding of the Yang-Yang action is close to that of \cite{lukYY}. Let us explain this point.
Suppose that the integrable model
depends on some parameters like, for example,  the radius of the cylinder in 
the case of Gibbs ensemble. Evaluating the Yang-Yang action on the solution to Bethe equation we obtain a function
which depend on these parameters only. 
We shall call this on-shell Yang-Yang action.

After this digression we present the relation between  $ \Theta\(
l, m| 0 \)$ 
evaluated at imaginary integer values on the one hand
and the on-shell Yang-Yang action on the other hand:
\begin{align}
 \Theta\(
ia, ib|0\)+\delta _{a,-b}
\pi \mathrm{sgn}(a)t_a(0)=\frac{\partial ^2}{\partial g_{a}\partial g_{b}}Y(\{g\})\,,
\quad a,b\in \mathbb{Z}/2\mathbb{Z}
\label{YY}
\end{align}
where the specialisation \eqref{sub} is implied in the right hand side after the derivatives 
are calculated. So, the one-point functions of the primary fields with $\al =2m+2n\frac{1-\nu}\nu$
and all their descendants are expressed as Hessians of the on-shell Yang-Yang action. 
This is similar to the Gaudin formula formally, but rather far from it in essence.

The paper is organised as follows. In Section \ref{fermions} we review 
the fermionic basis. In Section \ref{shG} we formulate our main conjecture concerning the one-point functions
in the shG case. We verify this conjecture against known results in Section \ref{comparing}. Finally, in 
Section \ref{further} a relation to the on-shell Yang-Yang action and to the representation
of the one-point functions of primary fields obtained by the method of separation of variables
is presented. 

\section{Fermionic basis}\label{fermions}

For absence of lattice formulation we shall
rely on our recent paper \cite{NS} where the fermionic
basis was defined as intrinsic property of the Liouville model, i.e. 
allows purely CFT definition. This is a basis
in the quotient space obtained by factorising out from the Verma module descendants
of the local integrals of motion. The defining property for the fermions is:
\begin{align}
&\sigma _1:\quad  \gammab^{\mathrm{CFT}*}_{2m-1}\longrightarrow u(\al)
\betab^{\mathrm{CFT}*}_{2m-1},\quad
\betab^{\mathrm{CFT}*}_{2m-1}\longrightarrow u^{-1}(-\al)\gammab^{\mathrm{CFT}*}_{2m-1}\,,
\label{reflection1}\\
&\sigma _2:\quad  \betab^{\mathrm{CFT}*}_{2m-1}\longrightarrow\gammab^{\mathrm{CFT}*}_{2m-1},\quad\quad\ \ \  \gammab^{\mathrm{CFT}*}_{2m-1}\longrightarrow\betab^{\mathrm{CFT}*}_{2m-1}\,,\nn
\end{align}
where
\begin{align}u(\al)=\frac{\nu\al  -(2m-1)(1-\nu)}{\nu\al  +(2m-1)}\,.\label{u}\end{align}
Let us mention one more property which clearly follows from \cite{NS}. There is
a duality $b\to b^{-1}$ which in our notations reads 
\begin{align}\nu\to\frac{\nu}{\nu-1}\,.\label{duality}\end{align}
We have
\begin{align}
\mathrm{duality:}\quad \betab^{\mathrm{CFT}*}_{2m-1}\longrightarrow\gammab^{\mathrm{CFT}*}_{2m-1},\quad\quad\ \ \  \gammab^{\mathrm{CFT}*}_{2m-1}\longrightarrow\betab^{\mathrm{CFT}*}_{2m-1}\,.\label{dualityCFTfermions}
\end{align}

For the second chirality we have to change $\al\to -\al$ in \eqref{u}.

The fermions are normalised as follows
\begin{align}
\betab^{\mathrm{CFT}*}_{I^+}
\gammab^{\mathrm{CFT}*}_{I^-}
\Phi_{\alpha}=C_{I^+,I^-}
\Bigl\{\mathbf{l}_{-2}^{n}
\ +\ \cdots\Bigr\}\Phi_{\alpha}\,,\quad \#(I^+)=\#(I^-)=n,\nn
\end{align}
where
$C_{I^+,I^-}$ is the Cauchy determinant cooked out of $1/(j^++j^--1)$. We have the same formula
for the second chirality. Explicit formulae up to the level 8 can be found in \cite{HGSIV,Boos},
in \cite{NS} a purely algebraic method is explained to obtain the
fermionic basis in general.

Now we eliminate the multipliers  in the formula  \eqref{reflection1}
defining the fermions
\begin{align}
& \betab^{*}_{2m-1}= D_{2m-1}(\al)\betab^{\mathrm{CFT}*}_{2m-1},\ \ \quad\qquad
  \gammab^{*}_{2m-1}= D_{2m-1}(2-\al)\gammab^{\mathrm{CFT}*}_{2m-1}\,,\nn\\
  & \bar\betab^{*}_{2m-1}= D_{2m-1}(2-\al)\bar\betab^{\mathrm{CFT}*}_{2m-1},\qquad
  \bar\gammab^{*}_{2m-1}= D_{2m-1}(\al)\bar\gammab^{\mathrm{CFT}*}_{2m-1}\,.\nn
\end{align}
In order to keep the duality, 
the constant $D_{2m-1}(\al)$ 
differs from the one used in the sG case \cite{HGSIV,OP,HGSV} 
by a ``CDD multiplier": a function periodic with period $4$. Namely, we set
\begin{align}
D_{2m-1}(\al)=\frac 1 {2\pi i}
\(\frac{\mub \Gamma (\nu)}{(\nu-1)^{\nu/2}}\)^{-\frac {2m-1}\nu}
\frac{\Gamma \(\frac \al 2 +\frac 1 {2\nu}(2j-1)\)
\Gamma \(\frac {2-\al} 2 +\frac {\nu-1} {2\nu}(2m-1)\)}{(m-1)!}\,.\nn
\end{align}
We introduced the dimensional multiplier in this definition in order to make
our fermions dimensionless.
The duality holds
provided  the following dimensional coupling constant
is self-dual:
$$\frac{\(\mub \Gamma (\nu)\)^{\frac 1 \nu}}{\sqrt{\nu -1}}\,.$$
This is indeed the case \cite{ZZ}, the simplest way to see it in the present
context consists in writing an explicitly self-dual quantity, the mass of the shG particle,
in terms of $\mub$ \cite{Alyoshascale}:
$$\frac{\(\mub \Gamma (\nu)\)^{\frac 1 \nu}}{\sqrt{\nu -1}}=\frac{m}{8\sqrt{\pi}}
\frac{\sqrt{\nu-1}}{\nu}
\ \Gamma \(\frac {\nu-1} {2\nu}\)
\Gamma \(\frac {1} {2\nu}\)\,.$$

Now we come to the relations \eqref{idm+}. For the moment we are unable to put them onto
solid ground of CFT, so, we proceed by analogy with the sG case. It can be shown that
due to the change of constants $D_{a}(\al)$ the constant in \eqref{idm+},
\eqref{comm} has to be modified, comparing to the sG case, to
\begin{align}
t_a(\al)=-\frac 1 {2\sin\frac{\pi}{\nu}(a+\nu\al)}\,.\nn
\end{align}
The constant $C_m(\al)$ is the same as  the one used in \cite{HGSV}:
\begin{align}
&C_m(\al)=\prod\limits_{j=0}^{m-1}C_1(\al+2j
{\textstyle \frac {1-\nu}\nu})\,,\label{Cm}\\
&C_1(\al)=(\mub \Gamma (\nu))^{4x}\  \frac{  \Gamma (-2\nu x) \Gamma (x)  \Gamma (1/2-x )}{\Gamma (2\nu x) \Gamma (-x) \Gamma (x+1/2 )}
 \,,\quad x=
{\textstyle \frac{\al}{2}+\frac{1-\nu}{2\nu}}\,.\nn
\end{align}
This constant   coincides with the ratio of the one-point functions of the shifted and unshifted primary
fields in infinite volume \cite{LZ}:
\begin{align}
C_m(\al)=\frac {\langle \Phi_{\alpha+2m\frac{1-\nu}{\nu}}(0)\rangle^{\mathrm{sG}}_{\infty}}
 {\langle \Phi_{\alpha}(0)\rangle^{\mathrm{sG}}_{\infty}}\,.\nn
\end{align}
In the paper \cite{HGSV} this was computed and served as an important support 
of entire construction. For the shG case
do not have such possibility for the moment, so, \eqref{Cm} is considered as a definition.

\section{One-point functions in sinh-Gordon model.} \label{shG}

We begin our  study of the shG model with brief review of 
its Euclidean version on a cylinder, our
exposition is based on the papers \cite{luksep,AlyoshshG}.
In  these papers the direction opposite to the one 
which is followed in \cite{HGSIV}  for  the sG model case is taken.
Namely,   the TBA-equation, which is very simple in the shG case,
is  chosen as starting point.
Then the $T$ and $Q$-functions related by Baxter equation 
are introduced via a series of
formal definitions. Finally, certain consistency check is performed.
Let us explain the reason for this change of point of view. 

Our basic object is the Euclidean field theory on a cylinder. For the sG (or, more precisely,
massive Thirring) model  the Euclidean field theory allows lattice
regularisation in the form of eight-vertex model.
Scaling behaviour near the point of the second order phase transition
can be observed for this model and scaling exponents can be computed \cite{Baxter}.
The paper  \cite{HGSV}  instead of
the eight-vertex model deals  with the inhomogeneous six-vertex model, which is an Euclidean
version of the construction of \cite{DDV}, but the motivation for that is rather technical than 
conceptual. 
The Matsubara transfer-matrices $T$ and $Q$ can be introduced as 
traces with respect to the two-dimensional and $q$-oscillator representations \cite{HGSIII},
exactly in the same way as for the continuous chiral CFT with $c<1$ \cite{BLZII}. 
In addition, working with the Matsubara transfer-matrices is a very reasonable choice for the sG case
because it leads to the Destri-DeVega equation \cite{DDV} instead of the horrifying
in the sG case system of TBA equations.

In the shG case the eight-vertex model should be 
replaces by a model with Boltzmann weights given by the universal R-matrix 
in the tensor product of two infinite-dimensional representations without highest weight. 
To the best of our knowledge 
the status of the phase transition for such lattice model
has not been clarified.
That is why we prefer not to rely on the lattice
construction of \cite{teschner,teschner1}. 
Since our only concern is the ground state in the Matsubara direction
we 
shall  follow the papers \cite{luksep,AlyoshshG} as has been said. 

The shG S-matrix is very simple, it gives rise to a single TBA equation
\begin{align}
\epsilon(\theta)=2\pi R m\cosh\theta -\int_{-\infty}^{\infty} {\log \(1+e^{-\epsilon(\theta')}\)}\Phi(\theta-\theta')d\theta'\,,\label{TBA}
\end{align}
where
$$\Phi (\theta)=\frac {1}{2\pi \cosh(\theta+\pi i\frac {\nu-2}{2\nu})}+
\frac {1}{2\pi \cosh(\theta-\pi i\frac {\nu-2}{2\nu})}=\int\limits _{-\infty}^{\infty}e^{ik\theta}\frac{\cosh\frac\pi{2\nu}(\nu -2)k}
{\cosh\frac\pi 2 k}\frac {dk}{2\pi}\,.$$
This is the basic equation, and the Matsubara data are defined using 
the pseudo-energy $\epsilon(\theta)$. Namely, define following \cite{luksep}
\begin{align}
\log Q(\theta)=-\frac {\pi Rm\cosh \theta}{\sin\(\frac{\pi}{\nu}\)}+\int\limits _{-\infty}^{\infty}
\frac {\log \(1+e^{-\epsilon(\theta')}\)}{\cosh(\theta -\theta ')}\frac {d\theta'}{2\pi}\,,\nn
\end{align}
here the growing at infinity term is chosen for consistency as explained below.
We have
\begin{align}
e^{-\epsilon(\theta)}=Q\Bigl(\theta +\frac{\pi i}{2\nu}(\nu -2)\Bigr)Q\Bigl(\theta -\frac{\pi i}{2\nu}(\nu -2)\Bigr)\,.\nn
\end{align}
From this equation one derives that $Q(\theta)$ satisfies the bilinear equation
\begin{align}
Q\Bigl(\theta +\frac{\pi i}{2}\Bigr)Q\Bigl(\theta -\frac{\pi i}{2}\Bigr)-
Q\Bigl(\theta +\frac{\pi i}{2\nu}(\nu -2)\Bigr)Q\Bigl(\theta -\frac{\pi i}{2\nu}(\nu -2)\Bigr)=1\,.
\label{qW}
\end{align}
Introduce
$\z=e^{\nu \theta}\,$.
It is easy to prove form \eqref{qW} that $T(\z)$ defined by the equation 
\begin{align}
T(\z)Q(\theta)=Q\Bigl(\theta +\pi i\frac{\nu-1}\nu\Bigr)+
Q\Bigl(\theta -\pi i\frac{\nu-1}\nu\Bigr)\,.\label{Baxter1}
\end{align}
is a single-valued function of $\z ^2$ with essential singularities at $\z=0,\infty$.
The functions $Q(\theta)$, $Q(\theta +\frac{\pi i }{\nu})$ can be considered as two different
solutions to the equation \eqref{Baxter1},  the left hand side of \eqref{qW} is their 
quantum Wronskian.

Certainly, while mounting all this construction
one has in mind defining {\it
a posteriori} the ground state eigenvalue of the Matsubara transfer-matrices which
is hard to define directly. In order to check that these definitions are reasonable,
in \cite{luksep} the behaviour of $T(\z)$ is investigated in the ultra-violet limit $R\to 0$ numerically.
It is shown that in this limit the asymptotics of $T(\z)$ for $\z\to 0,\infty$ correctly 
reproduce the eigenvalues of CFT integrals of motion with exactly the same
normalisation as in the sG case \cite{BLZII}. This is a very convincing argument.

Now by analogy with the sG case we want to deform
the kernel $\Phi(\theta)$ introducing $\Phi _\al(\theta)$ (we require $\Phi _0(\theta)=\Phi (\theta)$). 
The  Fourier image $\widehat{\Phi}(k,\al)$ should satisfy
the symmetry conditions
\begin{align}
\widehat{\Phi}(k,\al+2)=\widehat{\Phi}(k,\al)\,,\quad
\widehat{\Phi}(k,-\al)=\widehat{\Phi}(-k,\al)\,,\label{symshG}
\end{align}
and additional relation
\begin{align}
\widehat{\Phi}(k,\al+2{\textstyle \frac {1-\nu}{\nu}})=\widehat{\Phi}(k+2i,\al)\,.\label{shiftPhi}
\end{align}
The importance of the latter requirement will be clear soon.
It is not hard to find a deformation with required properties:
\begin{align}
&\Phi _\al(\theta)=
\frac {e^{i\pi\al}}{2\pi \cosh(\theta+\pi i\frac {\nu-2}{2\nu})}+
\frac {e^{-i\pi\al}}{2\pi \cosh(\theta-\pi i\frac {\nu-2}{2\nu})}=
\int\limits _{-\infty}^{\infty}e^{ik\theta}\widehat{\Phi}(k,\al)\frac{dk}{2\pi}\,,\nn\\
&
\widehat{\Phi}(k,\al)=\frac{ \cosh\pi\(\frac{\nu -2}{2\nu}k-i\al\)}
{\cosh\frac\pi 2 k}\,.\nn
\end{align}
Notice that contrary to $\widehat{R}(\theta,\al)$
which plays the same role in the sG case
\cite{HGSV}, the kernel $\widehat{\Phi}(k,\al)$,
as a function of $k$, does not have poles whose positions depend on $\al$. This simplification
is responsible for much simpler analytical properties of the one-point functions in $\al$. 
We would like to acknowledge that the idea of deforming 
the TBA-like equation with this kind of kernels in application to  one-point functions
appeared for the first time in \cite{Boosetal}.

By analogy with the sG case  we introduce the dressed resolvent which satisfies the equation
\begin{align}R_{\mathrm{dress},\al}-\Phi _\al\ast R_{\mathrm{dress},\al}=\Phi _\al\,,
\label{RdrshG}\end{align}
where
$$f\ast g=\int f(\theta)g(\theta)dm(\theta),\qquad dm(\theta)=\frac{d\theta}{1+e^{\epsilon(\theta)}}\,.$$
Define further
\begin{align}R_{\mathrm{dress},\al}(\theta,\theta')-\Phi _\al(\theta-\theta')=
\int\int \frac{dl}{2\pi}\frac{dm}{2\pi}\widehat{\Phi}(l,\al)
 \Theta(l,m|\al)
\widehat{\Phi}(m,-\al)e^{i(l\theta+m\theta ')}\,.\nn
\end{align}
The function $ \Theta(l,m|\al)$  satisfies an
equation, similar to (7.7) of \cite{HGSV}:
\begin{align}
 \Theta(l,m|\al)-G(l+m)-\int G(l-k)\widehat{\Phi}(k,\al)
 \Theta(k,m|\al)\frac{dk}{2\pi}=0\,.\label{eqThetashG}
\end{align}
where 
$$G(k)=\int e^{-ik\theta}\frac{d\theta}{1+e^{\epsilon(\theta)}}\,.$$
For the ground state the function $\epsilon(\theta)$ is even. 
So, one easily derives from \eqref{symshG}:
\begin{align}
& \Theta(l,m|-\al)= \Theta(m,l|\al)\,,\qquad
 \Theta(l,m|\al+2)= \Theta(l,m|\al)\,.\label{symTheta}
\end{align}
Now comes the relation \eqref{shiftPhi}. Solving \eqref{eqThetashG} by iterations and
shifting the integration contours one finds, similarly to the sG case,
\begin{align}
&\Theta(l,m|\alpha+2{\textstyle \frac{1-\nu}{\nu}})
\label{shiftshG}\\&
= \Theta(l+2i,m-2i|\alpha)-\frac{ \Theta(l+2i,-i|\alpha)
 \Theta(i,m-2i|\alpha)}
{ \Theta(i,-i|\alpha)
+\frac \pi {2\sin\pi(\frac{1}{\nu}+\alpha)}}\,.\nn
\end{align}
We shall soon see the importance of this property.

Changing in \eqref{shiftshG} $\al$ to $-\al$ and applying \eqref{symTheta} one easily finds
similar relation for shift in other direction, i.e. by $-2{\textstyle \frac{1-\nu}{\nu}}$.

\vskip .2cm
\noindent{\bf Main conjecture.} {\it We conjecture that similarly to the sG case the
one-point functions in fermionic basis are given by the determinant formula:
\begin{align}
&\frac {\langle
\betab^* _{I^+}\bar{\betab} ^*_{\bar{I}^+}
\bar{\gammab }^*_{\bar{I}^-}\gammab^*_{I^-}
\Phi_{\al}(0)
 \rangle_R}
{\langle\Phi_{\al }(0) \rangle_R}=
\mathcal{D} \(I^+\cup(- \bar{I}^+)\ |\ I^-\cup (-\bar{I}^-)|\al\)\,,\label{themainshG}
\end{align}
where for $A=\{a_j\}_{j=1,\cdots,n}$, $B=\{b_j\}_{j=1,\cdots,n}$ we set
\begin{align}
&\mathcal{D} (A|B|\al)=\prod\limits
_{j=1}^n \mathrm{sgn}(a_j) \mathrm{sgn}(b_j)\nn\\
&\times\frac 1 {\pi^n}\det\left.
\(  \Theta\(
 ia_j, ib_k|\al\)
-\pi \mathrm{sgn}(a_j)\delta _{a_j,-b_k}  t_{a_j}(\al)\)\right|_{j,k=1,\cdots ,n}\,.\nn
\end{align}
}
\vskip .2cm

Notice that in the infinite volume the formulae for the one-point functions of the primary fields and 
their Virasoro descendants coincide with the analytical continuation with respect to $\nu$ of sG ones. 

The relations \eqref{idm+} impose certain consistency requirements. Like in the sG
case all of them follow from the property \eqref{shiftshG} which explains its importance and hence
the necessity of the requirement \eqref{shiftPhi} imposed on the deformed kernel. Let us 
consider the simplest example. The relations \eqref{idm+} imply in particular
\begin{align}
\betab^*_1\gammab^*_1\Phi _{\al +2 \frac {1-\nu}\nu}=
\frac{C_1(\al)}{t_1(\al)}
\betab^*_3\betab_1\betab^*_1\bar\gammab^*_1\Phi_\al=
-C_1(\al)\betab^*_3\bar\gammab^*_1\Phi_\al\,.\nn
\end{align}
Hence
\begin{align}\frac{\langle\betab^*_1\gammab^*_1\Phi _{\al +2 \frac {1-\nu}\nu}(0)\rangle  _R}
{\langle\Phi _{\al +2 \frac {1-\nu}\nu}(0)\rangle  _R}\cdot\frac 
{\langle\Phi _{\al +2 \frac {1-\nu}\nu}(0)\rangle  _R}{\langle\Phi _{\al }(0)\rangle  _R}
=-C_1(\al)
\frac{\langle\betab^*_3\bar\gammab^*_1\Phi_\al(0)\rangle  _R}
{\langle\Phi_\al(0)\rangle  _R}
\,.\label{xx}
\end{align}
Notice that
$$\frac 
{\langle\Phi _{\al +2 \frac {1-\nu}\nu}(0)\rangle  _R}{\langle\Phi _{\al }(0)\rangle  _R}
=\frac{C_1(\al)}{t_1(\al)}\frac{\langle\betab^*_1\bar\gammab^*_1\Phi_\al(0)\rangle  _R}
{\langle\Phi_\al(0)\rangle  _R}\,.
$$
Computing all the ratios of
one-point functions by  the formula \eqref{themainshG} one observes that the identity \eqref{xx}
follows from \eqref{shiftshG} specialised at  $l=m=i$.

The rest of consistency relations follow from \eqref{shiftshG} by simple combinatorics \cite{HGSV}. 

\section{Comparing with known results}\label{comparing}

\subsection{Expectation values of energy-momentum tensor}

The eigenvalues of the local integrals of motion in Matsubara direction can be
obtained from the asymptotics of $\log T(\z)$ for $\z\to\infty$, $\z\to 0$ \cite{luksep,LZPDE}.
We do not go into details giving only the final formulae.
Define
\begin{align}
J_{2j-1}=
\frac{\pi^2mR }{\sin \frac \pi{\nu}}\cdot\delta_{2j-1,\pm 1}
-
 \int\limits _{-\infty}^{\infty} {\log \(1+e^{-\epsilon(\theta)}\)}e^{(2j-1)\theta}d
\theta\,,\quad  j\in\mathbb{Z}\,.\label{defJ}
\end{align}
Then for $2j-1>0$ we have 
\begin{align}
C_{2j-1}I_{2j-1}=J_{2j-1}\,,\qquad C_{2j-1}\bar I_{2j-1}=J_{-2j+1}\,, \nn
\end{align}
where 
$$C_{2j-1}=-\frac{\sqrt{\nu-1}}{\nu}\cdot\frac {\Gamma \(\frac{\nu-1}{\nu}(2j-1)\)\Gamma \(\frac{1}{\nu}(2j-1)\)}
{2\sqrt{\pi}j!}\(\frac{\mub\Gamma(\nu)}{(\nu-1)^{\nu/2}}\)^{-\frac {2j-1}\nu}
\,,\nn 
$$
and in the conformal limit the local integral $I_{2j-1}$ is such that its density starts with $:T(z)^{2j}:$
like in \cite{BLZII}. Notice that the  formula for $C_{2j-1}$ is self-dual.
The expectation values of components of the energy-momentum tensor $T$, $\bar{T}$, $\Theta$
can be expresses in terms of the Matsubara ground state energy $E(R)=I_1+\bar{I}_1$.
Repeating the computations of Section 10.3 of \cite{HGSV} one finds that our formulae
specialised to the cases $\betab^*_1\gammab^*_1$, $\bar\betab^*_1\bar\gammab^*_1$,
$\betab^*_1\bar\gammab^*_1$ give the correct result. This fixes the normalisation.

The authors of \cite{FFLZZ} observed an interesting relation between the vacuum expectation
values in sG model:
$$\langle T\bar T \rangle^\mathrm{sG}_\infty=-\(\langle \Theta \rangle^\mathrm{sG}_\infty\)^2\,.$$
In the paper \cite{zamTT} this relation was promoted to the identity
$$\langle T\bar T \rangle_R=\langle T\rangle\langle \bar T \rangle_R-\(\langle \Theta \rangle_R\)^2\,,$$
which holds for expectation values on a cylinder for any two-dimensional quantum field theory.
Proceeding with the same
computation as in  \cite{HGSV} we find agreement with this, the only {\it a priori} known,
determinant formula.

\subsection{LeClair-Mussardo formula} 
Define
\begin{align}F(\al)=\frac  {\langle \Phi _\al(0)\rangle_R}
 {\langle \Phi _\al(0)\rangle_\infty}\,.\nn\end{align}
 This function is periodic with respect to $\al$:
 \begin{align}
 F(\al+2)=F(\al)\,. \label{periodic}
 \end{align}
 Our fermionic formula gives
 \begin{align}
 \frac{F(\al+\frac {2}\nu)}{F(\al)}=1+\frac {2\sin\pi\(\al +\frac 1 \nu\)}{\pi}
\(e_1\ast e_{-1}+e_1\ast R_{\mathrm{dress},\al}\ast e_{-1}\)\,.\label{F/Fmy}
\end{align}
 We want to compare the first three terms of the large $R$ expansion with 
LeClair and Mussardo formula 
\cite{LM}. Let us write this formula explicitly for two reasons: first, it is convenient for us
to apply the duality transformation $b\to 1/b$ to the original formula \cite{LM}, second,
the three-fold integral is written with some typos in the paper \cite{LM}  which, nevertheless, can be easily
corrected using the general procedure described there. 

In order to simplify the comparison we introduce the notations:
$$2k=\nu\al,\quad [m]=\frac {\sin\frac{\pi}{\nu}m}{\sin\frac{\pi}{\nu}}\,.$$
Leclair and Mussardo claim the large $R$ expansion
\begin{align}
 F(\al)=1+\sum\limits
 _{n=1}^{\infty}\int \prod_{i<j\le n}\Phi (\theta _i-\theta _j)F_j(\theta _1,\cdots,\theta _n)\prod
 _{j=1}^n dm(\theta_j)\,.\nn
\end{align}
giving explicitly  the first three terms:
\begin{align}
	F_1 &= \frac {2\sin\frac{\pi}{\nu}} \pi[k]^2 \,,	\label{LMF}
	\\	
	F_2 &= \frac {2\sin\frac{\pi}{\nu}} \pi [k]^2  \left([k]^2c_{12}-\frac{[k-1][k+1]}{c_{12}}\right)\,,	\nonumber
	\\	
	F_3 &= \frac{[k]}{12}\bigg( A + B\big(c_{12}^2 + c_{23}^2 + c_{13}^2\big) + \frac{C}{c_{12}c_{23}c_{13}} + D\frac{c_{1}^2+c_{2}^2+c_{3}^2}{c_{12}c_{23}c_{13}} \bigg)\,,	\nonumber
\end{align}
where $c_{12}=\cosh(\theta _1-\theta_2)$, $c_1=\cosh (2\theta _1-\theta _2-\theta _3)$,
{\it etc},
\begin{align}
	A &=- 28[k-1][k][k+1]\Big( [k]^2+1\Big)\nn\\&+8\Big([k-2][k]^2[k+1]^2+[k-1]^2[k]^2[k+2]\Big)\nn
	\\& - 2\Big([k-2][k-1][k+1]^3+[k-1]^3[k+1][k+2]+[k-2][k]^3[k+2]-[k]^5\Big)\,,\nn
	\\
	B &= 8[k]^5\,,	\nonumber
	\\	
	C &= [k]^5 + 5[k-1][k]^3[k+1]+2[k-1]^2[k][k+1]^2+[k-2][k]^2[k+1]^2\nn\\&+[k-1]^2[k]^2[k+2]	
	-[k-2][k-1][k+1]^3-[k-1]^3[k+1][k+2]\nn\\&-[k-2][k]^3[k+2]-3[k-2][k-1][k][k+1][k+2]\,,\nn
		\\
	D &=-4[k-1][k]^3[k+1]\,.	\nonumber
\end{align}
Computing the same kind of large $R$ expansion for  $F(\al+\frac {2\pi}{\nu})/F(\al)$ 
we find that the first three terms have the same structure as in \eqref{LMF} with slightly simpler coefficients:
\begin{align}
	\widetilde A &=\frac{[2k+1][2]}{6}\Big( 3 [2k+1][2k-1][2] - [2k+1][2k]([2]^2+6) -2 [4k]\Big)\,,	
	\nonumber
	\\
	\widetilde B &= \frac{2}{3}[2k+1]^3\,,	\nonumber
	\\
	\widetilde C  &= \frac{[2k+1][2]}{24} \Big[ [4k]\big(3[2]^2-4\big)
	-2[2k+1][2k]\big([2]^2-6\big)  \Big]\,,	\nonumber
	\\
	\widetilde D &= -\frac{1}{6}[2k+1]^2[2k][2]	\,.\nonumber
\end{align}
Now it is easy to see that this coincides with our formula \eqref{F/Fmy} for the first
two iterations of $R_{\mathrm{dress},\al}$:
$$\frac{F(\al+\frac {2}\nu)}{F(\al)}=1+\frac {2\sin\pi\(\al +\frac 1 \nu\)}{\pi}
\(e_1\ast e_{-1}+e_1\ast \Phi _\al\ast e_{-1}+e_1\ast \Phi _\al\ast \Phi _\al\ast e_{-1}+\cdots\)\,.$$

\subsection{Classical case} 
The role of Planck constant is played by $b^2$ (recall $\nu=1+b^2$).
In \cite{luksep} Lukyanov gives the classical approximation
for $F(\al)$
(our $\al$ is twice Lukyanov's $\al$). It is convenient for our goals to present his result in the form:
\begin{align}
\log F(\al)=\frac 1 {b^2}\int\limits _0^\al d\al \int\limits _{-\infty}^{\infty}\frac{d\theta}{2\pi i}
\log\(\frac {1-e^{-r\cosh\theta-\pi i \al}}{1-e^{-r\cosh\theta+\pi i \al}}
\)+O(b^0)\,,\label{lukclass}
\end{align}
where $r=2\pi m R$.
This formula is obtained in \cite{luksep} in two different ways: by applying the steepest descent method to the integral obtained by the separation of variables (see \eqref{lukint} below), 
and by evaluating the classical action on the solution to the sG equation with a puncture. 
So, the formula \eqref{lukclass} serves Lukyanov to verify the answer obtained by the
separation of variables method. We have to check that our conjecture agrees with this important formula. 

From \eqref{lukclass} we get
\begin{align}
& \frac{F(\al+2\frac {1-\nu}{\nu})}{F(\al)}=\exp\(
- \frac 1 {\pi i}\int\limits _{-\infty}^{\infty}
\log\(\frac {1-e^{-r\cosh\theta-\pi i \al}}{1-e^{-r\cosh\theta+\pi i \al}}\)
d\theta\)+O(b)\label{F/Fluk}
\end{align}

On the other hand from \eqref{periodic}, \eqref{F/Fmy} we find
\begin{align}
 \frac{F(\al+2\frac {1-\nu}{\nu})}{F(\al)}= 1-\frac {2\sin\pi\al }{\pi}e_{1}\ast^\mathrm{cl} E_{-1}+O(b)
,\nn
\end{align}
where  the function $E_{-1}$ satisfies the equation:
\begin{align}
E_{-1}=e_{-1}+\Phi _\al ^\mathrm{cl}\ \ast ^\mathrm{cl} \ E_{-1}\,.\label{classeqF}
\end{align}
The last formulae contain the classical limits $\Phi _\al ^\mathrm{cl}$ and $\ast ^\mathrm{cl} $. Let us 
compute them.

We immediately find
$$\Phi ^{\mathrm{cl}}_\al(\theta)=\frac {e^{-\pi i \al}}{2\pi i \sinh(\theta -i0)}-\frac {e^{\pi i \al}}{2\pi i \sinh(\theta +i0)}\,.
$$
This implies, in particular,
$$\Phi ^{\mathrm{cl}}_0(\theta)= \delta(\theta)\,.$$
Using the latter  one solves explicitly the equation  \eqref{TBA} obtaining
$$1+e^{\epsilon ^{\mathrm{cl}} (\theta)}= e^{r\cosh \theta}\,.$$
So, the limit of $f\ast g$ is
$$f\ast^\mathrm{cl} g =  \int\limits _{-\infty}^{\infty} f(\theta)g(\theta)e^{-r\cosh\theta}d\theta\,.$$

Introduce the function
$$G(\theta)=\int\limits _{-\infty}^{\infty} \frac {E_{-1}(\theta ')}{2\pi \cosh(\theta -\theta ')}e^{-r\cosh(\theta ')}d\theta '\,.$$
Obviously,
\begin{align}
e^{-r\cosh \theta}E_{-1}(\theta)=G(\theta+{\textstyle \frac {\pi i} 2})+G(\theta-{\textstyle \frac {\pi i} 2})\,.
\label{F=G+G}
\end{align}
The equation \eqref{classeqF} turns into a simple boundary problem for $G(\theta)$:
\begin{align}
G(\theta+{\textstyle \frac {\pi i} 2})\bigl(1-e^{-\pi i \al-r\cosh\theta}\bigr)
+G(\theta-{\textstyle \frac {\pi i} 2})\bigl(1-e^{\pi i \al-r\cosh\theta}\bigr)=e^{-\theta-r\cosh\theta}\,.\label{eqG}
\end{align}
We solve this equation, and then find $E_{-1}(\theta)$ using \eqref{F=G+G}.

Introduce the notation
\begin{align}H(\theta)=\frac {1-e^{-r\cosh\theta-\pi i \al}}{1-e^{-r\cosh\theta+\pi i \al}}-1\,,
\label{defH}
\end{align}
and two functions
$$X_{\pm}(\theta)=\exp\Bigl(-\frac 1 {2\pi i}\int\limits _{-\infty}^{\infty} \frac {e^{\theta -\theta '}}{\sinh (\theta -\theta '\pm i0)}
\log\(1+H(\theta ')\)d\theta '\Bigr)\,.$$
Then after some simple computations we come to the conclusion that the equality of
\eqref{F/Fluk} and \eqref{F/Fmy} is equivalent to the identity
\begin{align}
&\exp\Bigl(-\frac 1 {\pi i}\int\limits _{-\infty}^{\infty} \log(1+H(\theta))d\theta\Bigr)=1-\frac 1 {2\pi i }\int\limits _{-\infty}^{\infty} \frac 
{(2+H(\theta))H(\theta )}{1+H(\theta)}d\theta\label{classicalidentity}\\
&
+\frac 1 {(2\pi i)^2}\int\limits _{-\infty}^{\infty}\int\limits _{-\infty}^{\infty} \frac{
1}{\sinh(\theta -\theta ')}
\(e^{\theta -\theta '}\frac {X_-(\theta)}{X_+(\theta ')}-
e^{\theta' -\theta }\frac {X_-(\theta')}{X_+(\theta )}\)H(\theta)H(\theta ') d\theta d\theta '\,.\nn
\end{align}
This identity holds for any rapidly decreasing at $\pm \infty$ function $H(\theta)$. The proof is given in Appendix.

From this computation we learn 
that  finding a quantum deformation of
reasonably simple classical formula \eqref{lukclass} is not straightforward. 
One has to consider the ratio \eqref{F/Fluk}, then rewrite it using the identity
\eqref{classicalidentity} which can be interpreted in terms of the  
integral equation \eqref{classeqF}. The latter  allows a quantum deformation.
It would be interesting to try to guess one-point functions for other integrable
models using similar procedure.

\section{Further remarks}\label{further}

\subsection{Yang-Yang action}
As has been said in the Introduction it is useful to consider
the  partition function for generalised Gibbs ensemble \cite{RDYO}
$$Z(\{g_{2j-1}\})=\Tr \(e^{-\sum_{p-\infty}^{\infty}g_{2p-1}\mathcal{I}_{2p-1}}\)\,,\qquad g_{2j-1}>0,\ \forall j\,,$$
where $\mathcal{I}_{2j-1}$ are space integrals of motion. The TBA equation is
derived by standard procedure  \cite{YY,AlTBA} introducing the
pseudo-energy $\epsilon(\theta)$ and minimising the free energy:
\begin{align}
\epsilon(\theta)=\sum_{j=-\infty}^{\infty}g_{2j-1}e^{(2j-1)\theta}
-\int_{-\infty}^{\infty} {\log \(1+e^{-\epsilon(\theta')}\)}\Phi(\theta-\theta')d\theta'\,.\label{genTBA}
\end{align}
We assume that the series $\sum_{j=-\infty}^{\infty}g_{2j-1}z^{2j-1}$ has infinite
radius of convergency, then the TBA equation is perfectly well-defined, the iterations
converge very fast. 
We define the $Q$-functions as above:
\begin{align}
\log Q(\theta)=-\sum_{j=-\infty}^{\infty}\frac{g_{2j-1}}{2\cos \pi\frac{(2j-1)(\nu -2)}{2\nu}}e^{(2j-1)\theta}+\int\limits _{-\infty}^{\infty}
\frac {d\theta'}{2\pi}\frac {\log \(1+e^{-\epsilon(\theta')}\)}{\cosh(\theta -\theta ')}\,.\label{Qgen}
\end{align}
The meaning of this function is not quite clear. It may be possible to define it
in the spirit of  \cite{KS}, starting from the lattice model \cite{teschner,teschner1}
on a cylinder and putting inhomogenieties into the Matsubara 
transfer-matrix. We expressed our
reservations concerning the lattice model 
at the beginning of Section \ref{shG}, but if  such construction were available
one would be able to  generalise formulae for the eigenvalues of the Matsubara local integrals:
\begin{align}
J_{2j-1}=
\frac{\pi g_{-(2j-1)}}{2\sin \pi\frac{|2j-1|}{\nu}}
-
 \int\limits _{-\infty}^{\infty} {\log \(1+e^{-\epsilon(\theta)}\)}e^{(2j-1)\theta}d
\theta\,,\quad  j\in\mathbb{Z}\,. \label{J}
\end{align}
For the moment we cannot rely on such hypothetical construction, so, \eqref{J}
is considered as a formal definition.

Set
$$R_{\mathrm{dress}}=R_{\mathrm{dress},0}\,.$$
It is easy to see that
$$\frac{\partial\epsilon}{\partial g_{2j-1}}=e_{2j-1}+R_{\mathrm{dress}}\ast e_{2j-1}\,,$$
This is used in order to evaluate the derivative of $J_{2j-1}$ with respect to parameters:
\begin{align}
&\frac{\partial}{\partial g_{2j-1}}J_{2k-1}=
\frac{\pi}{2\sin \pi\frac{|2j-1|}{\nu}}\delta _{2j-1,-(2k-1)}
+
e_{2j-1}\ast e_{2k-1}+e_{2j-1}\ast R_{\mathrm{dress}}\ast e_{2k-1}\,.\nn
\end{align}
The resolvent $R_{\mathrm{dress}}$ is symmetric which ensures the
existence of potential 
$$J_{2k-1}=\frac{\partial Y(\{g\})}{\partial g_{2k-1}}\,.
$$
We call the potential $Y(\{g\})$ the on-shell Yang-Yang action
because in the case of usual Gibbs ensemble
it  coincides with the one defined in \cite{lukYY}.
Comparing with the previous formulae we find the formula announced in the Introduction \eqref{YY}:
\begin{align}
 \Theta((2j-1)i,(2k-1)i|0)+\delta _{2j-1,-(2k-1)}
\frac {\pi}{2\sin\frac{|2j-1|}{\nu}}=\frac{\partial^2 Y(\{g\})}{\partial g_{2j-1}\partial g_{2k-1}}\,.\nn
\end{align}
Hence the expectation values of all the fermionic descendants of $\Phi_0=1$ are expressed as
Hessians of the on-shell Yang-Yang action. It may look as a restriction that we are able to
write such a nice formula only for the fermionic descendants of $1$, but these
descendants include the Virasoro descendants of all the primary fields
$\Phi_{2n\frac{1-\nu}{\nu}}$ $n\in\mathbb{Z}$ due to \eqref{idm+}. In other words, provided 
$ \Theta((2j-1)i,(2k-1)i|0)$ are given we {\it define} 
$ \Theta((2j-1)i,(2k-1)i|2n\frac{1-\nu}{\nu})
$ 
inductively using \eqref{shiftshG}.The resulting formula for
$ \Theta((2j-1)i,(2k-1)i|2n\frac{1-\nu}\nu)$ can be rewritten 
as $(n+1)\times (n+1)$ determinant containing $ \Theta((2j-1)i,(2k-1)i|0)$.
Clearly, $ \Theta((2j-1)i,(2k-1)i|2n\frac{1-\nu}\nu)$ obtained in this
way must coincide with the values of the analytical function 
$ \Theta(l,m|2n\frac{1-\nu}\nu)$ satisfying the integral equation
\eqref{eqThetashG}. 
Together with the requirement of $2$-periodicity of
the one-point 
functions at radius $R$ normalised to
the one-point
functions at $R=\infty$,  this provides their values at $\al=2p+2q\frac{1-\nu}{\nu}$. For
irrational $\nu$ these points are dense on $\mathbb{R}$, and generic $\al$ appears just by
continuity. The consistency is guaranteed, as usual, by \eqref{shiftshG}. 

For the Gibbs ensemble
the on-shell Yang-Yang action was shown in \cite{lukYY} to coincide with
the action of the classical Euclidean shG model with special boundary conditions. 
It would be very interesting to generalise the construction of \cite{lukYY} to
the case of generalised Gibbs ensemble. This may give an
interesting interpretation of the one-point functions.

\subsection{Expectation value as integral over separated variables}
We would like to finish this paper with one more remark. It is known that the matrix 
elements of quantum integrable models related to the affine algebra $U_q(\slth)$
can be expressed in terms of rather nice integrals \cite{toda,sgsep} using the separated variables 
representation \cite{sklyanin}. 
Lukyanov observed \cite{luksep} that the best
application of this method is the shG model. 
Actually the separated variables were originally introduced by Sklyanin exactly in
order to treat the shG model which is out of reach of the Bethe Ansatz \cite{sklyaninsinh}. 
In order to write the formula of \cite{luksep} in manifestly self-dual form it is convenient to
introduce
$$\tilde{\nu}=1+b^{-2}\,,$$
such that
$$\frac 1 \nu+\frac 1 {\tilde{\nu}}=1\,.$$
Then the expectation value of the primary
field $\Phi _\al(0)$ is given by properly regularised integral of the form:
\begin{align}
\langle \Phi _\al(0)\rangle _R=\int
\prod _{j=-\infty}^{\infty}d\theta _j
\prod _{j=-\infty}^{\infty}Q^2(\theta _j)e^{(\tilde{\nu}+\nu)\al \theta _j}
\prod _{i<j}\sinh \nu(\theta _i-\theta _j)\sinh 
 \tilde{\nu}(\theta _i-\theta _j)\,,\label{lukint}
\end{align}
all integrals are taken from $-\infty$ to $\infty$.
The integral \eqref{lukint} can be thought about as a generalisation of matrix integrals in which
instead of square of one Wandermonde determinant a product of two of them, including
differently the coupling constant, is inserted. So, one of results of this paper consists in
proposing an exact answer for this rather complicated integral. More precisely, we consider the
ratio of expectation values for primary with $\al$ shifted by $2\frac {1-\nu}\nu$ , but due to
analyticity one can  solve the simple difference equation for logarithm. 

We believe that our formulae hold for the case of generalised Gibbs ensemble. This
means that these generalised matrix integrals can be evaluated for a set
of functions $ Q(\theta )$ parametrised by $g_{2j-1}$. One has to be careful at this point
because we required $g_{2j-1}>0$ for the sake of the equation \eqref{genTBA}, but
in the equation \eqref{Qgen} they come divided by $\cos \pi\frac{(2j-1)(\nu -2)}{2\nu}$.
For generic $\nu$ these cosines do not vanish, but they may turn negative which is
bad for the convergence of integrals in \eqref{lukint}. In every particular case the
behaviour of $ Q(\theta )$ for $\theta \to\pm\infty$ should be investigated, and it might
be necessary to move the contours of integration in \eqref{lukint} from the real axis to the
rays on which the integrals converge. 

One more generalisation is possible. Actually the only property of $Q(\theta)$ which we
really need for our construction is the quantum Wronskian relation \eqref{qW}. We can allow
function $Q(\theta)$ to have zeros in the strip $-\pi<\mathrm{Im}(\theta)<\pi$. In the
case of the usual Gibbs ensemble this 
corresponds to considering excited states for Matrubara transfer-matrix. The TBA equation
changes in this case \cite{AlyoshshG} in rather simple way. The most general situation
which we can imagine includes the function $Q(\theta)$ with zeros in the strip, and
with the asymptotics corresponding to the generalised Gibbs ensemble \eqref{Qgen}. 
It would be interesting to find other applications of the generalised matrix model integral
\eqref{lukint}.

\section{Appendix}
In this Appendix we prove the identity \eqref{classicalidentity}.
The function $X_{+} (\theta )$ 
allows analytical continuation to the strip $0<\mathrm{Im}(\theta)<\pi$, it coincides with
$X_{-} (\theta )$ on the upper bank of this strip. We have
\begin{align}
&X^+(\theta)=(1+H(\theta))X^-(\theta)\,,\nn\\
&X^+(\infty)=\exp\Bigl(-\frac 1 {\pi i}\int\limits _{-\infty}^{\infty} \log(1+H(\theta))d\theta\Bigr)\,,\quad X^+(-\infty)=1\,.
\nn
\end{align}
Consider the two-fold integral in in \eqref{classicalidentity}. For
definiteness let us understand the denominator as $\sinh (\theta -\theta '+i0)$. 
Then, changing the integration variables we rewrite:
\begin{align}
I_2=\frac 1 {2(\pi i)^2}\int\limits _{-\infty}^{\infty}\int\limits _{-\infty}^{\infty} \frac{
e^{\theta -\theta '}}{\sinh(\theta -\theta '+i0)}
\frac {X_-(\theta)}{X_+(\theta ')}H(\theta)H(\theta ') d\theta d\theta '
+\frac 1 {2\pi i}\int\limits _{-\infty}^{\infty}\frac {H(\theta)^2}
{1+H(\theta )} d\theta \,.\nn
\end{align}
Evaluating the integral with respect to $\theta$
\begin{align}
&\int\limits_{-\infty}^{\infty}\frac {e^{\theta-\theta'}X_-(\theta)H(\theta)}{2\pi i\sinh(\theta -\theta '+i0)}d\theta
=\int\limits_{-\infty}^{\infty}\frac {e^{\theta-\theta'}(X_+(\theta)-X_-(\theta))}{2\pi i\sinh(\theta -\theta '+i0)}d\theta\nn\\&=
\int\limits_{-\infty}^{\infty}\frac {e^{\theta-\theta'}X_+(\theta)}{2\pi i\sinh(\theta -\theta '+i0)}d\theta
-\int\limits_{-\infty}^{\infty}\frac {e^{\theta-\theta'}X_-(\theta)}{2\pi i\sinh(\theta -\theta '-i0)}d\theta\nn+X_-(\theta ')\\&=
-\exp\Bigl(-\frac 1 {\pi i}\int\limits _{-\infty}^{\infty} \log(1+H(\theta))d\theta\Bigr)+X_-(\theta ')\,,\nn
\end{align}
we obtain
\begin{align}
I_2=-\frac 1 {\pi i}\exp\Bigl(-\frac 1 {\pi i}\int\limits _{-\infty}^{\infty} \log(1+H(\theta))d\theta\Bigr)\int\limits _{-\infty}^{\infty} \frac{H(\theta)}{X_+(\theta)}d\theta 
+\frac 1 {2\pi i}\int\limits _{-\infty}^{\infty}\frac {2+H(\theta)}
{1+H(\theta )}H(\theta) d\theta \,.\nn
\end{align}
It remains to compute:
\begin{align}
\frac 1 {\pi i}\int\limits _{-\infty}^{\infty} \frac{H(\theta)}{X_+(\theta)}d\theta =
\frac 1 {\pi i}\int\limits _{-\infty}^{\infty}\(\frac{1}{X_-(\theta)}- \frac{1}{X_+(\theta)}\)d\theta=
\exp\Bigl(\frac 1 {\pi i}\int\limits _{-\infty}^{\infty} \log(1+H(\theta))d\theta\Bigr)-1\nn\,,
\end{align}
this finishes the proof.

{\it Acknowledgements.}\quad 
We are grateful to O. Babelon, M. Jimbo, L. Takhtajan and J.-B. Zuber for discussions. Special
thanks are due to A. Its and S. Lukyanov whose advices helped to improve considerably our paper.  

\noindent
Research of SN is supported by Universit\`a Italo Francese
grant ``Vinci".
Research of FS is supported  by DIADEMS program (ANR) contract number BLAN012004.      



\end{document}